\def\G1915{GRS $1915$+$105$}
\def\X1550{XTE J$1550$--$564$}
\def\J1655{GRO J$1655$--$40$}
\def\eg{{\it e.g.} }
\def\etal{{\em et al. }}
\def\ie{{\em i.e. } }

\documentclass[]{aa}

\usepackage{epsfig}
\usepackage{epsf}

\titlerunning{Timing Analysis of \G1915}
\authorrunning{Rodriguez \etal }
\begin{document}

\title{Energy dependence of a Low Frequency QPO in GRS 1915+105}

\author{J. Rodriguez\inst{1} \and Ph. Durouchoux\inst{1}  \and I. F. Mirabel\inst{1,}\inst{2} \and Y. 
Ueda\inst{3}  \and M. Tagger\inst{1} \and K. Yamaoka\inst{4}}
\offprints{J. Rodriguez \\
(rodrigue@discovery.saclay.cea.fr)}

\institute{DSM/DAPNIA/Service d'Astrophysique (CNRS URA 2052), CEA Saclay, 91191 Gif-sur-Yvette, France,\and Instituto de Astronom\'{\i}a y F\'{\i}sica del Espacio/CONICET, Buenos Aires, Argentina, \and Institute of Space and Astronautical Science, Yoshinodai 3-1-1, Sagamihara, Kanagawa 229-8510, Japan, \and RIKEN The Institute of Physical and Chemical Research, Hirosawa 2-1, Wako, Saitama 351-0198, Japan.}
\date{Received date; Accepted date}

\abstract{We analyze a set of three RXTE Target of Opportunity observations of
 the Galactic microquasar \G1915, observed on April 2000, during a 
multi-wavelength campaign. 
 During the three observations, a strong, variable low frequency ($2-9$ Hz) 
quasi periodic oscillation (hereafter QPO), often referred to as the 
ubiquitous QPO,  is detected together with its first harmonic.\\
We study the spectral properties of both features, and show that :
1) their frequency variations are better correlated with the soft X-ray flux
 ($2-5$ keV), favoring thus the location of the QPO in the accretion disk;
2) the QPO affects more the hard X-rays, 
usually taken as the signature of an inverse compton scattering of the soft 
photons in a corona; 
3) the fundamental and its harmonic do not behave in the same manner: 
the fundamental sees its power increase with the energy  up to 40 keV, 
whereas the harmonic increases up to $\sim 10$ keV.\\
The results presented here could find an explanation in the context of the
 Accretion-Ejection Instability, which could appear as a rotating spiral
 or hot point located in the disk, between its innermost edge and the 
corotation radius. 
The presence of the harmonic could then be a signature of the
 non-linear behavior of the instability.
The high-energy ($>40$ keV) decrease of the fundamental would favor
 an interpretation where most or all of the quasi-periodic modulation
 at high energies comes, not from the comptonized corona as usually
 assumed, but from a hot point in the optically thick disk.\\
}

\maketitle

\section{Introduction}
X-ray binaries  exhibit strong 
X-ray emission, from the soft ($\sim 0.1$ $keV$) to the hard X-rays (up to a 
few hundred keV), sometimes up to the MeV domain. The emission processes are 
thought to occur in the close 
vicinity of a stellar-mass compact object (either a Neutron Star or a Black 
Hole), the soft part of the spectrum being  usually taken as the thermal 
emission of an 
accretion disk, whereas the hard part is thought to be the manifestation of an
 inverse compton
 scattering of the soft photons, with relativistic electrons present in a 
hot coronal medium. 
The sources may be distinguished by several characteristics, such 
as the companion mass, whenever this latter is known, the shape of their 
spectra, or by the presence of strong collimated ejecta.
In the latter case, the similarity with AGN led to the definition of 
microquasars (Mirabel \etal, 1992), some of them 
known to be sources with superluminal jets (Mirabel \& Rodr\'{\i}guez, 1999).\\
\G1915 has first been discovered as a Soft X-ray Transient by WATCH on board 
{\itshape{GRANAT}}
(Castro-Tirado \etal, 1992), and then identified as the first Galactic source
to have ejections  with apparent superluminal motion (Mirabel \& Rodr\'{\i}guez, 1994). 
The distance to the source has been estimated to 12.5 kpc, its inclination 
$\sim 70^{\circ}$, and the velocity of the jet 0.92c (Mirabel \& Rodr\'{\i}guez, 1994). 
Since then, the source has been observed with many X-ray satellites, and
 its spectrum is typical of that of Black Hole Candidates (BHC), such as 
\J1655. Only recently, however, the spectral type of the companion has been 
identified as a K--M III star (Greiner \etal, 2001), classifying the source 
as a low mass X-ray binary. The mass of the primary has been estimated to
$14\pm4$ $M_{\odot}$ (Greiner, Cuby \& McCaughrean, 2001), confirming the 
black hole nature of the compact object.\\
With the launch of the Rossi X-ray Timing Experiment (RXTE), and the excellent
 timing capacities of both its pointed instruments, the {\itshape 
{Proportional Counter Array}} (PCA) and the {\itshape{High Energy X-ray Timing
 Experiment}} (HEXTE), many X-ray Binaries and \G1915 in particular, have been
 discovered to exhibit Quasi Periodic Oscillations (QPOs), in several 
ranges of frequency (a few mHz up to hundred, and kilohertz in the case of
neutron star primary). Though no physical explanation has yet been widely
accepted, the QPOs are thought to occur in the close vicinity of the 
compact object.\\
 Furthermore, it has been pointed out by Psaltis \etal \ (1999), that the QPOs 
could represent the same type of variability in both neutron stars and black 
hole systems, constraining the theoretical models, and giving important 
clues to the physics of these phenomena. 
In particular the study of QPOs should give important informations on the 
accretion flow, and thus on the physics of the disk.\\
The detection of several types of QPOs can be  attributed to
different mechanisms, depending in particular on the source spectral state.\\
 We will only focus here on the strong $\sim 0.5-10$ $Hz$ QPO, present during 
the low/hard spectral state of \G1915, often called ``ubiquitous'', since 
it is nearly always
present in that state and often observed in other Black Hole Binaries 
(e.g. \X1550, or \J1655).
In that case, several authors have pointed out correlations between 
the frequency of the oscillations and some of the spectral 
parameters, such as the flux (Swank \etal, 1997; Markwardt \etal, 1999), the 
temperature of the disk (Muno \etal, 1999), and 
the disk color radius (Rodriguez \etal, 2002).\\
All these correlations constrained the location of the QPO in or close to 
the disk, and the systematic study of the QPO parameters should lead 
to a better understanding of the accretion and ejection mechanisms, thought 
to occur in this region.\\
Recently a new mechanism has been proposed by Tagger \& Pellat \ (1999), to 
extract energy and angular momentum from the inner regions of the disk 
(permitting, thus the accretion) and transport them toward the corotation 
radius of the spiral wave formed in the disk, where they can be emitted 
directly toward the corona 
(Tagger \& Pellat, 1999; Varni\`ere \& Tagger, 2001).\\
It has been shown by Rodriguez \etal \ (2001), and Varni\`ere \etal \ (2002)
, that this model could explain  
the different frequency vs. radius correlations observed in \J1655 compared to
 \G1915 or (as had been found by Sobczak \etal, 1999) \X1550.\\
This model could also explain the correlations found by Mirabel \etal (1998), 
Eikenberry \etal (1998), Ueda \etal (2002, our observations being part of this
 latter work)
 during the $\sim 30$ min cycle (Tagger, 1999 for a possible scenario), between
 X-ray light curves and the infrared and radio emissions, considered  as the 
synchrotron signatures of an expanding ejected blob of material, relating then
 the energy needed to accelerate those blobs, 
to the one extracted from the accretion.\\

\begin{table*}[htbp]
\centering
\begin{tabular}{|c|c|c|c|c|c|c|}
\hline
Date & MJD & Obs Id & Interval $\#$ & Time start (UT)& Time stop (UT) &PCUs ``On''\\
\hline
\hline
04 17 2000 & $51651$ & $50405-01-01-00$ & $1$ & $12h52m15s$& $13h42m55s$ & $0-4$ \\
$ $ & $ $ &  & $2$ &$14h27m43s$ & $15h18m39s$ &$0-4$\\
\hline
04 22 2000 & $51656$ & $50405-01-02-00$ & $1$ & $09h21m35s$ & $10h15m27s$ &$0-4$\\
$ $ &  & $50405-01-02-01$ & $2$ & $10h55m59s$ & $11h38m07s$ &$0,2-4$\\
$ $ &  & $50405-01-02-02$ & $3$ & $12h31m59s$ & $13h14m07s$ &$0,2-4$\\
\hline
04 23 2000 & $51657$ & $50405-01-03-00$ & $1$ & $07h40m31s$ & $08h35m27s$&$0,2,3$\\
$ $ & & $ $ & $2$ & $09h16m15s$ & $09h59m59s$&$0,2,3$\\
\hline
\end{tabular}
\caption{List of the Observations reduced; the interval time are those define 
by the PCA good time intervals as defined in section \ref{sec:data}. Relative time zero corresponds to $12h52m15s$, start of the good time interval for 
interval \# 1.}
\label{tab:ref}
\end{table*}

We present here observations of the source taken as a RXTE Target of 
Opportunity, in April 2000.
In section \ref{sec:data} we present the data reduction and analysis methods 
used; in section \ref{sec:17th}, we examine  the first of the three 
observations, which is the most variable one, and focus then on the dynamical
 properties of 
the source, observed in different energy ranges. In section \ref{sec:22-23} 
we  study the data of the following observations, where the source is much 
more steady, and thus, more
 adapted to extract the QPO parameters with high accuracy; we will interpret 
our observations in the last part of this paper.

\section{Data reduction and analysis}
\label{sec:data}

The source has been observed on April $17^{th}$, $22^{nd}$ and $23^{rd}$, 2000
as a target of opportunity.
We have reduced and analyzed the processed data 
using the  FTOOLS package (update 5.04). 
Observations IDs, exact time intervals, and dates 
are shown in table \ref{tab:ref}.\\
We first extracted, for the three observations, lightcurves covering the entire
 PCA energy range, from binned data with $2^{-7}$ s = $7.8125$ ms resolution, 
and event data with $2^{-16}$ s = $15.25878$ $\mu$s, which were rebinned 
during the extraction process to $7.8125$ ms.\\
In all cases, lightcurves were extracted from all the PCUs that were
simultaneously turned ``on'' over a single interval (\ie 5 on Apr. $17^{th}$,
 and $22^{nd}$ first interval, four during the two following intervals 
that day, and three on Apr. $23^{rd}$). We combined all PCUs and all layers 
 to get the most possible incoming flux. The exact PCA configuration over 
each interval is given in table \ref{tab:ref}.\\
``Good Time Intervals'' (GTIs) were defined when the 
elevation angle was above $10^{\circ}$, the offset pointing less than 
$0.02^{\circ}$, and  we also excluded the data taken while crossing the SAA.
\\
Background lightcurves were generated using the PCABACKEST tool, from 
standard2 data, and subtracted from the raw lightcurves. 
We then generated power spectra and dynamical power spectra (hereafter DPS)
using POWSPEC 1.0, calculating each FFT over $\sim4$ s time intervals (2048 
bins in each intervals), and averaging then the result over 4 intervals. 
The resultant DPS has a resultant time bin $\sim 16$ s, comparable to the time
resolution of the standard 2 lightcurves.
To follow the evolution of the QPOs parameters with the energy, we 
 extracted, in the same standard way, lightcurves in five PCA energy channels:
 absolute channel $0-11$ (in Matrix epoch4 corresponding to $<2-4.99$ keV),
 channel $12-29$ ($4.99-12.68$ keV), channel $30-46$ ($12.68-20.06$ keV),
channel $47-89$ ($20.06-39.29$ keV), channel $90-174$ ($39.29-80.04$ keV).
We then produced DPS and power spectra, as explained above, in each energy 
range.
\begin{figure}[htbp]
\epsfig{file=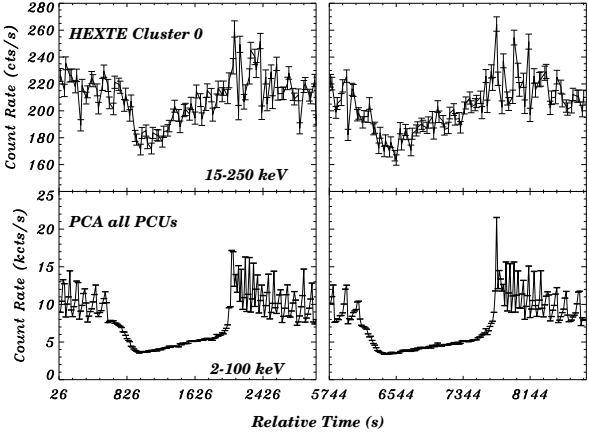,width=\columnwidth}
\caption{The source on April $17^{th}$. Standard lightcurves covering in both 
cases the entire instrument energy range ($\sim 15-250$ $keV$ for HEXTE, and 
$\sim 2-100$ $keV$ for PCA); {{\itshape{Upper Panel : }}} HEXTE Cluster 0
 standard lightcurves with 16 s time bins, {{\itshape{Lower Panel : }}}
PCA standard 2 lightcurves with 16 s time bins. X axis is in unit of s; Y axis 
is in unit of cts/s (upper panels), and kcts/s (lower ones). 
Error bars are $1\sigma$ statistical errors.}
\label{fig:bothlite}
\end{figure}

\section{First Observation : on April $17^{th}$}
\label{sec:17th}
We extracted from both instruments standard lightcurves with 16 s 
time resolution, using the standard PCA and HEXTE reduction steps, for this 
observation; they are plotted on figure \ref{fig:bothlite}.\\

\begin{figure}[htbp]
\epsfig{file=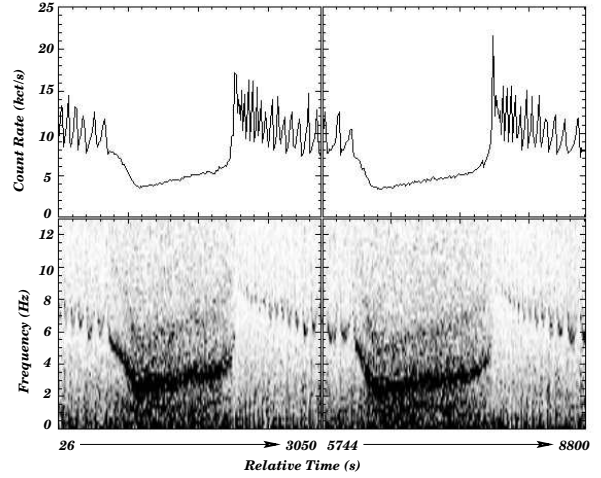,width=\columnwidth}
\caption{PCA Lightcurve of the source during the 
whole observation; {{\itshape {Lower Panel : }}} Dynamical power spectra of 
the source on April $17^{th}$; the gap in the data corresponds to occultation 
due to the orbit. X axis is the relative time (time 0 is defined in table \ref{tab:ref}), in units of s. Y axis is in kcts/s (upper panels), and in Hz (lower ones).}
\label{fig:dynpow17-04}
\end{figure}

\begin{figure}
\epsfig{file=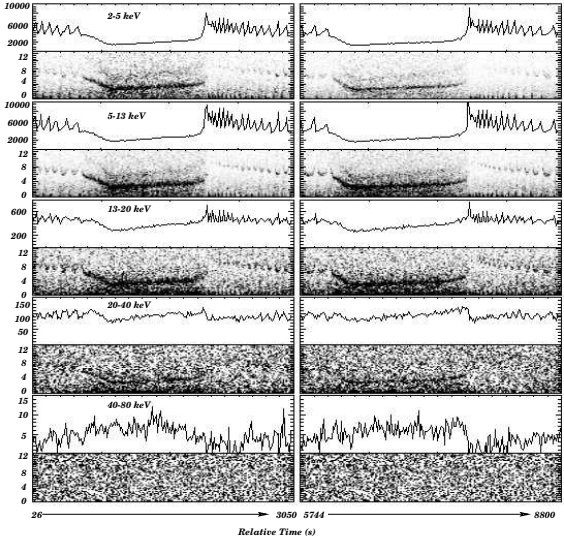,width=\columnwidth}
\caption{Dynamical power spectra of the source on April $17^{th}$ in 
the five PCA energy ranges; X axis in all cases are the time, covering relative
 time from 24 s to 3050 s (Left), and relative time from 5744 s to 8800 s (Right); upper panels are the standard 2 lightcurves in the indicated energy ranges. Y axis are in units of kcts/s for the upper panels
, and in units of Hz for the lower ones.}
\label{fig:rangedynpow}
\end{figure}

The source is in a {{\itshape{$\alpha$ state }}} as defined by Belloni \etal 
(2000).
PCA dynamical power spectra, covering the entire PCA energy range ($\sim2-100$
 $keV$), are shown 
on figure \ref{fig:dynpow17-04} together
 with the PCA lightcurves.\\
The source presents large flux variations on short
time scales ($\sim 100$ s), together with a single QPO whose frequency has a
similar behavior (figure \ref{fig:dynpow17-04}). Then around time $\sim 600$ s
 (first interval), and $\sim 6000$ s  (second), a 
large $\sim 1000$s dip occurs (figure \ref{fig:bothlite}). During that  time, 
the QPO frequency varies from $9$ Hz to $2.25$ Hz, and a strong
 second QPO appears with a frequency $\sim$ twice that of the fundamental, 
following
the same frequency variations (figure \ref{fig:dynpow17-04}).
Then around relative time 2032 s (first interval), and 7716 s (second interval)
, a sudden and large soft X-ray spike, reaching 
$\sim 4.8 \times$ (respectively $\sim 6.4 \times$) the dip minimum flux, for 
the first (respectively second) interval, occurs and the source returns to a 
state similar to the one before the dip. Here the harmonic disappears, while 
the fundamental returns to a larger frequency and behaves as before the dip.
In addition we show in fig. \ref{fig:rangedynpow}, DPS in the five energy 
ranges defined in section \ref{sec:data}, together with the 
corresponding lightcurves. One can immediately see
that above 20 keV the harmonic is  absent or very faint, and that above 40 keV 
(probably due to the high noise) the QPO disappears.
We also see on figure \ref{fig:rangedynpow} the evolution of the flux 
variations with the energy; the large dip seems to be smoothed with the energy.
\\
We extracted from the soft lightcurves the relative time and the value of the
flux of the peak occuring just before the dip (relative time $554$ s, for 
the first interval, and $6064$, for the second one); we then re-did the same 
procedure for the minimum of the dip (relative time $954$ s for the first
 interval, and $6368$ s for the second one), and we thus could estimate 
the relative amplitude of the variation of the flux, at the time where, also, 
the fundamental QPO sees its frequency varying from 9 to 2.25 Hz. 
We did this in each energy range, at the same times 
(allowing a maximum of two bins ($\sim \pm 32$ s) of difference between each 
range). Results are shown in table \ref{table:var}.

\begin{table*}[htbp]
\centering
\begin{tabular}{c c c}
 Energy Range (keV)& Variation Rate Interval \# $1$ ($\%$)&  Variation Rate Interval \# $2$ ($\%$)\\
\hline
\hline
 $2$--$5$ & $72.95 \pm 0.63$ & $71.19 \pm 0.66$\\
\hline
 $5$--$13$ & $70.24 \pm 0.61$ & $69.61 \pm 0.63$\\
\hline
 $13$--$20$ & $47.84 \pm 1.78$ & $42.37 \pm 1.88$\\
\hline
$20$--$40$ & $22.18 \pm 4.26$ & $21.29 \pm 4.07$\\
\hline
\end{tabular}
\caption{Variations of the flux with the energy, between the last peak before the dip, and the bottom of the dip for the two intervals of April 17.}
\label{table:var}
\end{table*}
Note that the soft spike corresponds in the higher energy range (above 20 keV) 
to a sudden 
decrease of the flux, indicating the cooling, or the disappearance of
 a part of the corona 
(multi-wavelength results can be found in Ueda \etal, 2002 ).
\section{Second and Third Observation : April $22^{nd}$ and $23^{rd}$}
\label{sec:22-23}
\begin{figure}[htbp]
\epsfig{file=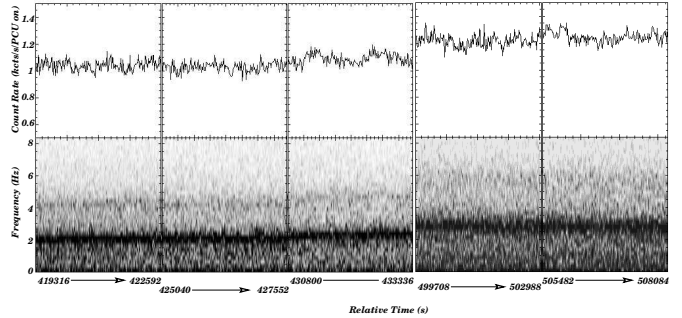,width=\columnwidth}
\caption{Plot of the 16s PCA lightcurves (upper panel), and the dynamical power spectra 
(bottom), covering the three good time intervals of April $22^{nd}$ 
(left Panel) and April $23^{rd}$ (Right Panel). Once again time zero is  
April $17^{th}$ good
 time interval start. Y axis of upper panels is in unit of kcts/s/PCUs on (see table \ref{tab:ref} for the PCA configuration over each interval), while that of lower panels is in Hz.}
\label{fig:22dynpow}
\end{figure}

As the lightcurves and dynamical power spectra did not present variations as
 strong as on the previous date, we did not focus here on the 
dynamical evolution of the QPO, but
we just tried to correlate the QPO parameters with the energy range.
Figure \ref{fig:22dynpow} shows the lightcurves with 
the
 dynamical power spectra from all the GTIs of both observations.
The source is in a 
{{\itshape {$\chi$ state}}} of Belloni \etal, 2000, characterized by a steady 
flux.

\begin{figure}
\epsfig{file=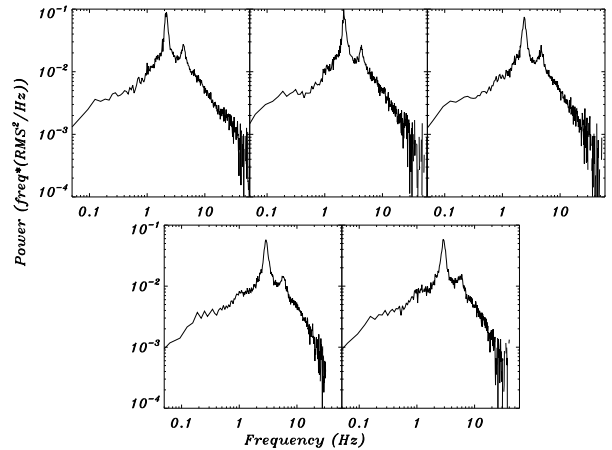,width=\columnwidth}
\caption{Power spectra of the source on April $22^{nd}$ (up), and 
April $23^{rd}$ (lower panel). Y axis is the power in terms of $Freq*RMS^2 / Hz$, while X axis are the frequencies.}
\label{fig:powspec22}
\end{figure}

Power spectra covering the entire PCA range,shown on figure 
\ref{fig:powspec22}, are fitted with a model consisting of two broad 
 lorentzians (continuum) , plus sharper ones, modeling
the QPO features. When the presence of the QPOs was not obvious, we estimated 
the parameters by freezing the lorentzian centroid frequency to the value
found in the other energy ranges, and allowing both the width and the power 
to vary. In the case of the $40-80$ keV range, since the statistics from 
single interval was poor, we choosed to merged the observations were the QPO 
frequency was found to be close, i.e. intervals $\#1$ and $\#2$ from April 22,
 and intervals $\#1$ and $\#2$ from April 23; interval $\#3$ from April 22 was 
fitted alone.
Results from the fits for all the 
energy ranges defined in section \ref{sec:data} are shown in table 
\ref{table:paramqpo}.
No variations similar to those of April 17 are present here; the flux remains
fairly constant around a mean value 1050 cts/s/PCU-on, for the April 22 two 
first intervals, rising slowly to $\sim 1100$ cts/s/PCU-on, for the April 22 
third interval, and reaching $\sim 1200$ cts/s/PCU-on, on April 23.
As expected, in the same time intervals the fundamental QPO sees its frequency
 slowly increase with time from
$\sim 2.14$ Hz (on Apr. 22) to $\sim 2.9$ Hz (on Apr. 23) (figure 
\ref{fig:22dynpow}, and table \ref{table:paramqpo}). The harmonic is still 
present during the five intervals, with a frequency varying from $\sim4.3$ Hz 
(on Apr. 22), to $\sim 5.8$ Hz on April 23 first interval.\\
We plotted in figure \ref{fig:correl} the evolution of the QPO power
vs. energy range for the five GTIs. 
The upper points represent the behavior of the
fundamental QPO, and the lower that of the harmonic; we can see that
the power of the fundamental increases  up to $40$ keV, and them seems to 
decrease, whereas that of the harmonic seems to peak between the
$5-13$ and $13-20$ keV ranges.
\begin{figure}[htbp]
\epsfig{file=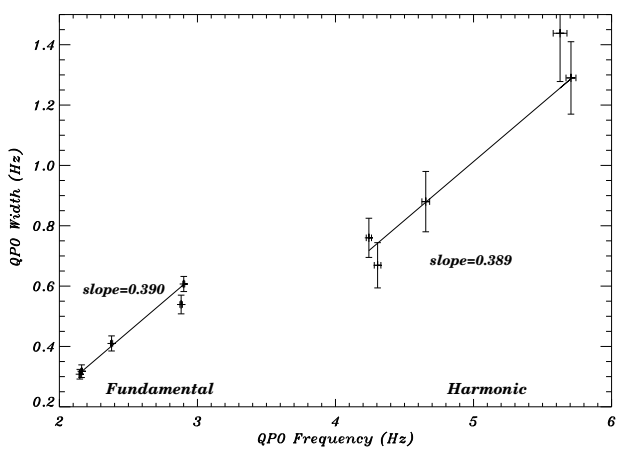,width=\columnwidth}
\caption{Plot of QPO FWHM vs. QPO Frequency, for the five intervals covering
Apr. $22^{nd}$ and $23^{rd}$. Both axis are in units of Hz. In the two case the solid lines represent the best fit. The slopes are indicated in each cases.}
\label{fig:width}
\end{figure}
Figure \ref{fig:width} represents the evolution of the QPOs width vs. 
their frequencies. Both distributions of points can be well fitted by lines 
of slopes $0.390$ for the fundamental, and $0.389$ for the harmonic. 
The zero abscissa values are found to be $-0.526052$ for the fundamental, 
and $-0.932459$ for the harmonic (although their physical meaning is not clear)
. It is clearly visible on the plot that both QPOs are tightly 
correlated, the width of the harmonic being $\sim$ twice that of the 
fundamental (resulting thus in a Q value ($=\frac{frequency}{FWHM}$) similar 
for both).
\section{Results and Interpretation}
\begin{figure}[htbp]
\epsfig{file=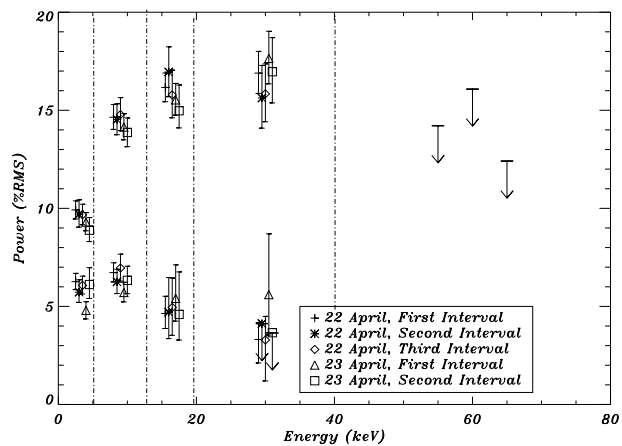,width=\columnwidth}
\caption{Plot of the Power vs. the energy range; each group of points
represents the value over the whole energy range delimited by the dash-dotted line. The upper group of points in each cases corresponds to the fundamental QPO, the lower being the harmonic. Error bars are  $1\sigma$ confidence level. Arrows are the $95\%$ upper limits. The three last arrows represent, respectively, the result from Apr. 22 intervals 1 \& 2 merged, Apr.22 interval 3, and Apr. 23 interval 1 \& 2 merged.}
\label{fig:correl}
\end{figure}
The April 17 observation confirms and expands the conclusion of
Markwardt \etal \ (1999) and Muno \etal \ (1999), that the QPO frequency is
better correlated with the soft flux, but seems stronger in the higher
energy bands (which is confirmed by the following dates).\\
In addition a precise study of the lightcurve of the same date shows
that the $\sim 30$ min dips are smoothed with the energy, and that the
sudden increase of the soft flux (the spike) is anti correlated with
the hard flux; indeed the spike, in both interval, corresponds to a
major decrease of the flux in the $20-40$ keV, and $40-80$ keV bands,
usually considered to be emitted by the corona. 
The soft spike marks here the transition from the low hard state (C 
state of Belloni \etal, 2000), to a soft high state (A-B states). Within the 
interpretation in terms of disk states, this transition and the rapid 
variations following (interpreted as rapid transitions through A B C states 
(Belloni \etal, 2000)) can be seen as a succession of rapid replenishments 
and  disappearances of the innermost parts of the disk (Belloni \etal, 1997). 
The behaviour of the corona may appear difficult to understand, 
since the abrupt cutoff of the hard X-rays could either  be the 
manifestation of a sudden cooling of the relativistic electrons by the 
re-emergence of a high soft flux, or the disappearance of the corona 
(by advection or ejection).\\
Thanks to a large number of multi-wavelength observations, the radio and
 infra red behaviors of \G1915 have now been widely studied for years. 
In particular, former studies such as the one presented in Mirabel \etal, 1998,
 or Eikenberry \etal,
 1998 had linked the soft X-ray spike (transition from low hard to soft high 
state) with radio and infra red flares. Dhawan \etal (2000) 
have shown that indeed superluminal ejections took place during abrupt change
 in the X-ray state of the source. More recently, Klein-Wolt \etal (2001) 
have found a strong correlation between radio events (radio oscillations, 
compact jets, large radio flares), and state C properties (duration, 
transition to other states). It is, however, to be noted that Klein-Wolt \etal
, did not find any simultaneous radio - alpha state observations.
Furthermore, ``The Largest Multi-wavelength Campaign'' on \G1915 presented in 
Ueda \etal (2002), shows that the state transitions on Apr. $17^{th}$ are
followed by radio flares consistent with an ejection of material starting at 
the state transition. This leads us to suggest that the abrupt cutoff 
of the hard X-rays is more probably related to the disappearance of a part of 
the corona, blown away under the form of a synchrotron emitting blob of 
material detected in the infra red, and radio domains 
(fig. 1 and 2 in Ueda \etal, 2002).\\
On the other hand, the behavior of the QPO and its harmonic at high 
 energies poses severe constraints on theoretical models. The 
 decrease of the QPO power above $40$ keV may indicate that not all the 
 corona is affected. The decrease of the harmonic above $\sim 20$ keV 
 also raises very challenging questions.\\ 
These could find an explanation in the context of the
 Accretion-Ejection Instability (Tagger and Pellat, 1999), which has
 been shown to form a rotating spiral structure in the disk, similar to
 galactic ones but driven by magnetic stresses rather than by
 self-gravity.  The spiral arms should be expected to heat as well as
 compress the gas in the disk, and thus to appear as a rotating spiral
 or hot spot.  The harmonic would then be a signature of the
 non-linear behavior of the spiral, just as the gas form shocks (and
 thus strong harmonics of the underlying 2-armed spiral) along galactic
 spiral arms.  The high-energy cutoff of the fundamental could, then, favor
 an interpretation where most or all of the quasi-periodic modulation
 at high energies comes, not from the comptonized corona as usually
 assumed, but from a hot point in the optically thick disk. This would
 be consistent with the previous result (Rodriguez \etal, 2001; Rodriguez 
\etal, 2002) that
 the anomalously small color radius of the disk, often observed in some
 Black-Hole Binaries, could actually be interpreted by the black-body
 emission of a small area hot point in the disk.
 We could in principle have an estimate of its physical size,
 by adding a blackbody model in the spectral fits (such as the {\itshape
{BBODYRAD}} model of XSPEC), one of the parameters being the normalized area
 of the emitting region, (since the black body luminosity is proportionnal to 
the area). But the limited sensitivity and spectral resolution of the present 
data do not allow any realistic fit. We expect that future instruments will 
provide better constraints on this problem .\\
It would be very tempting to consider the width of the QPO as a measure
 of the size (due for example to the differential rotation acting between 
the inner and outer edges of the spot). But the fact that we are dealing with a
 QPO probably rules out this explanation, since it has to result from a 
quasi-stationary feature in the disk. This is precisely the case for the AEI,
 where a standing spiral wave results in a quasi-stationary feature rotating
 at a single frequency. In this context the width of the QPO would correspond
 to the coherence time of this pattern, fixed either by non-linear effects or
 by variations in the background disk equilibrium, {\em e.g.} the inner disk 
radius or other disk parameters (temperature, magnetization, etc.)..\\
The spot physical properties (\eg its temperature) may also depend on a number
 of external parameters, hard to deduce from the observations, such as 
the $\beta$ ratio (the ratio between thermal and magnetic pressure), 
which drives the instability (see for example Varni\`ere \etal, 2002, for a 
discussion on the effects of this parameter), or even the efficiency of the 
instability. Indeed, in a non linear regime for example, the amount of 
energy deposited in the disk (under the form of shocks) would be much greater, 
and would locally warm it up much more than in the linear case.\\
Further observational and theoretical work should, however, allow to test this
 hypothesis: by producing, from numerical simulations of the
 instability (such as Caunt and Tagger, 2001), synthetic light curves
 of the QPO, and by fitting the observed energy dependence of the
 modulated light curve by a high-temperature, hotter black body over a
 small area of the disk rather than the usual power-law of the coronal
 emission.

\begin{acknowledgements}
The authors would like to thank S. Corbel, M. Muno, P. Varni\`ere, T.Foglizzo, 
 and  the anonymous referee for usefull discussions and comments which allowed 
to improve the quality of the paper.\\
IFM acknowledges partial support from Fundac\'{\i}on Antorchas.\\
We also thank the {{\itshape {Athena help}}} at GSFC for appreciable
help on the RXTE data reduction processes.\\
This research has made use of data obtain through the High Energy
Astrophysics Science Archive Center Online Service, provided by the
NASA/Goddard Space Flight Center.
\end{acknowledgements}

\bibliographystyle{plain}

\newpage
\begin{table*}[htbp]
%\centering
\hspace{-1cm}
\begin{tabular}{|c|c|c|c|c|c|c|c|c|c|c|}
\hline
Date & $\#$ & Energy range (keV) & $f_{QPO1} (Hz)$ & $Q_1$ & $\% RMS_1$ & $f_{QPO2} (Hz)$ & $Q_2$ & $\% RMS_2$ & $\chi^2$ (d.o.f.) \\
\hline
\hline
$04$ $22$ $2000$ & $1$ & PCA$^*$ & $2.148_{-0.006}^{+0.006}$ & $6.97$ & $12.47_{-0.57}^{+0.62}$ & $4.242_{-0.019}^{+0.019}$ & $5.58$ & $6.31_{-0.42}^{+0.48}$ & $69.13 (62)$ \\
  & & $2-5$ keV & $2.137_{-0.006}^{+0.008}$ & $7.12$ & $9.92_{-0.46}^{+0.46}$ & $4.270_{-0.02}^{+0.022}$ & $5.34$ & $6.26_{-0.4}^{+0.42}$ & $89.2 (62)$\\
 & & $5-13$ keV & $2.142_{-0.005}^{+0.007}$ & $6.88$ & $14.64_{-0.61}^{+0.65}$ & $4.249_{-0.018}^{+0.019}$ & $6.85$ & $6.72_{-0.46}^{+0.50}$ & $67.62 (62)$\\
 & & $13-20$ keV &$2.151_{-0.006}^{+0.007}$ & $6.68$ & $16.16_{-0.73}^{+0.73}$ & $4.270_{-0.038}^{+0.038}$ & $7.89$ & $4.63_{-0.76}^{+0.87}$ & $62.11 (62)$\\
& & $20-40$ keV & $2.143_{-0.008}^{+0.01}$ & $6.00$ & $16.89_{-1.03}^{+1.10}$ &$4.27$ {\itshape{frozen}} &$>6.1$ &$<4.17$ & $30.19 (36)$\\
\hline
 & $2$ & PCA & $2.161_{-0.006}^{+0.007}$ & $6.79$ & $12.27_{-0.75}^{+0.78}$ & $4.305_{-0.024}^{+0.024}$ & $6.43$ & $5.39_{-0.51}^{+0.57}$ & $102.5 (62)$\\
 & & $2-5$ keV & $2.152_{-0.007}^{+0.008}$ & $7.24$ & $9.70_{-0.66}^{+0.72}$ & $4.326_{-0.025}^{+0.025}$ & $6.21$ & $5.72_{-0.52}^{+0.63}$ & $108.6 (62)$\\
 & & $5-13$ keV & $2.162_{-0.007}^{+0.006}$ & $6.90$ & $14.53_{-0.78}^{+0.79}$ & $4.286_{-0.023}^{+0.024}$ & $7.10$ & $6.24_{-0.59}^{+0.64}$ & $80.10 (62)$\\
 & & $13-20$ keV & $2.169_{-0.009}^{0.008}$ & $5.60$ & $16.95_{-1.26}^{+1.28}$ & $4.333_{-0.092}^{+0.102}$ & $5.82$ & $4.71_{-1.35}^{+1.75}$ & $87.52 (42)$\\
 & & $20-40$ keV & $2.182_{-0.012}^{+0.012}$ & $6.28$ & $15.62_{-1.53}^{+1.66}$ & $4.30$ {\itshape{frozen}} & $>6.14$ &$<4.12$ & $46.24 (47)$\\
 & $1-2$ Merged &$40-80$ keV& $2.15$ {\itshape{Frozen}}&$>15$&$<14.21$&&&&$7.69 (10)$\\
\hline
 & $3$ & PCA & $2.378_{-0.008}^{+0.008}$ & $5.8$ & $12.31_{-0.68}^{+0.70}$ & $4.654_{-0.028}^{+0.029}$ & $5.28$ & $5.94_{-0.50}^{+0.55}$ & $87.98 (62)$\\
 & & $2-5$ keV & $2.361_{-0.008}^{+0.009}$ & $5.84$ & $9.67_{-0.5}^{+0.53}$ & $4.690_{-0.025}^{+0.026}$ & $5.39$ & $6.07_{-0.43}^{+0.46}$ & $94.8 (62)$\\
 & & $5-13$ keV & $2.382_{-0.007}^{+0.008}$ & $5.71$ & $14.78_{-0.81}^{+0.86}$ & $4.691_{-0.031}^{+0.03}$ & $5.54$ & $6.95_{-0.66}^{+0.7}$ & $70.54 (62)$\\
 & & $13-20$ keV & $2.391_{-0.009}^{+0.009}$ & $5.97$ & $15.78_{-1.16}^{+1.26}$ & $4.558_{-0.1}^{+0.127}$ & $6.16$ & $4.94_{-1.42}^{+1.49}$ & $100.1 (62)$\\
 & & $20-40$ keV & $2.376_{-0.011}^{+0.013}$ & $5.78$ & $15.83_{-1.42}^{+1.52}$ &$4.65$ {\itshape{Frozen}} & $>5.8$ & $<4.48$ & $43.58 (36)$\\
  & &$40-80$ keV  & $2.35$ {\itshape{Frozen}}& $>10.21$ & $<16.08$ &&&&$19.29 (28)$\\
\hline
\hline
$04$ $23$ $2000$ & $1$ & PCA & $2.901_{-0.007}^{+0.009}$ & $4.77$ & $12.07_{-0.47}^{+0.51}$ & $5.706_{-0.036}^{+0.036}$ & $4.42$ & $5.23_{-0.35}^{+0.43}$ & $87.98 (62)$\\
& & $2-5$ keV & $2.871_{-0.009}^{+0.009}$ & $5.31$ & $9.30_{- 0.44}^{+0.48}$ & $5.832_{-0.039}^{+0.041}$ & $5.12$ & $4.80_{-0.41}^{+0.44}$ & $83.09 (62)$\\
& & $5-13$ keV & $2.905_{-0.007}^{+0.008}$ & $5.12$ & $14.14_{-0.65}^{+0.68}$ & $5.752_{-0.033}^{+0.034}$ & $5.70$ & $5.71_{-0.48}^{+0.52}$ & $79.32 (62)$\\
& & $13-20$ keV & $2.921_{0.009}^{+0.01}$ & $5.35$ & $15.53_{-0.78}^{+0.84}$ & $5.535_{-0.146}^{+0.132}$ & $4.22$ & $5.48_{-1.16}^{+1.71}$ & $122.8 (62)$\\
 & & $20-40$ keV & $2.925_{-0.015}^{+0.015}$ & $4.91$ & $17.64_{-1.29}^{+1.38}$ &$5.940_{-0.16}^{+0.2}$ &$7.36$ &$5.60_{-2.06}^{+3.10}$ & $76.89 (59)$\\
\hline
& $2$ & PCA & $2.882_{-0.008}^{+0.01}$ & $5.34$ & $11.47_{-0.54}^{+0.64}$ & $5.627_{-0.051}^{+0.051}$ & $3.91$ & $5.41_{-0.47}^{+0.50}$ & $100.1 (62)$\\
& & $2-5$ keV & $2.866_{-0.012}^{+0.012}$ & $5.83$ & $8.88_{-0.57}^{+0.64}$ & $5.714_{-0.065}^{+0.067}$ & $3.19$ &$6.11_{-0.71}^{+0.85}$ & $79.77 (62)$\\
& & $5-13$ keV & $2.883_{-0.009}^{+0.01}$ & $5.49$ & $13.86_{-0.72}^{+0.74}$ & $5.640_{-0.063}^{+0.062}$& $3.90$ & $6.33_{-0.68}^{+0.72}$ & $94.63 (62)$\\
& & $13-20$ keV & $2.899_{-0.01}^{+0.013}$ & $5.71$ & $14.97_{-0.87}^{+1.31}$ & $5.65$ {\itshape{Frozen}} & $4.92$ & $4.58_{-1.3}^{+2.16}$ & $63.84 (42)$\\
& & $20-40$ keV & $2.901_{-0.019}^{+0.019}$ & $4.98$ & $16.96_{-1.59}^{+1.73}$ &$5.65$ {\itshape{Frozen}}& $>17.65$ &$<3.65$ & $36.86 (29)$\\
 & $1-2$ Merged & $40-80$ keV & $2.85$ {\itshape{Frozen}} & $>7.5$ & $<12.41$ &&&&$50.04 (41)$\\
\hline
\end{tabular}
\caption{results of the fittings for the three observations, for all the energy
 ranges defined in section \ref{sec:data}.
$^*$Instrument entire energy range ($\sim2-100$ keV).}
\label{table:paramqpo}
\end{table*}

\end{document}